\documentclass[twocolumn]{aastex631}

\usepackage{newtxtext,newtxmath}
\usepackage{bm}

\begin{document}
\bibliographystyle{apj}
\title{Turbulence and Star Formation Suppression in Elliptical Galaxies: The Role of Active Galactic Nucleus Jet-Wind Interaction}

\author[0009-0009-7849-2643]{Minhang Guo}
\affiliation{Astrophysics Division, Shanghai Astronomical Observatory, Chinese Academy of Sciences, 80 Nandan Road, Shanghai 200030, P.R.China}
\affiliation{ShanghaiTech University, 393 Middle Huaxia Road, Shanghai 201210, P.R.China}
\affiliation{University of Chinese Academy of Sciences, 19A Yuquan Road, Beijing 100049, P.R.China}

\author[0000-0001-9658-0588]{Suoqing Ji}
\affiliation{Center for Astronomy and Astrophysics and Department of Physics, Fudan University, Shanghai 200438, P.R.China}
\affiliation{Key Laboratory of Nuclear Physics and Ion-Beam Application (MOE), Fudan University, Shanghai 200433, P.R.China}

\author[0000-0003-3564-6437]{Feng Yuan}
\affiliation{Center for Astronomy and Astrophysics and Department of Physics, Fudan University, Shanghai 200438, P.R.China}

\author[0000-0003-0900-4481]{Bocheng Zhu}
\affiliation{National Astronomical Observatories, Chinese Academy of Sciences, 20A Datun Road, Beijing 100101, P.R.China}

\correspondingauthor{Suoqing Ji, Feng Yuan}
\email{sqji@fudan.edu.cn, fyuan@fudan.edu.cn}

\newcommand \mdot{\dot M} 
\newcommand \rg{r_{g}}

\date{}


\begin{abstract}
Winds and jets are symbiotic when the accretion rate is low, according to black hole accretion theory. Both components are potentially important for active galactic nucleus (AGN) feedback, but previous works typically include only jets with free parameters. We perform hydrodynamical simulations of an isolated elliptical galaxy with both jets and winds included. The key features discriminating our simulations from others are that our simulations resolve the Bondi radius for reliable black hole accretion rate calculation and use parameters from GRMHD simulations. By selectively activating jets and winds, we examine their individual and combined effects. We find that effective AGN feedback, which is capable of generating strong turbulence and subsequently increasing central gas entropy and suppressing cool gas condensation and star formation, occurs only when both jets and winds operate simultaneously. The physical mechanism is the interaction between winds and jets: this interaction produces strong shear at their interface, leading to turbulence via the Kelvin-Helmholtz instability. In contrast, neither jets nor winds alone can generate strong turbulence due to the insufficient shear. The turbulence produced by wind-jet interaction is predominantly solenoidal in nature, giving rise to a broad energy spectrum approximately following a Kolmogorov-like power law and a dissipation rate $\sim 10^{-27}\,\mathrm{erg\,cm^{-3}\,s^{-1}}$ in the interstellar medium, consistent with observations. 
Our findings highlight the importance of simultaneously considering both jets and winds in studying the effects of AGN feedback in the evolution of elliptical galaxies.
\end{abstract}

\keywords{galaxies: evolution --- ISM: structure --- methods: numerical --- hydrodynamics
}

\section{Introduction}
\label{sec:intro}

While the mechanical power of active galactic nucleus (AGN) jets, inferred from bubble cavity enthalpies, appears sufficient to balance radiative cooling, the precise energy transfer mechanisms remain insufficiently understood. Observationally, indirect methods provide evidence for turbulent processes: X-ray surface brightness fluctuations \citep{2014Zhuravleva} and the broadening of emission lines \citep{2025XRISM} reveal turbulence in the intracluster medium (ICM) and circumgalactic (CGM) as revealed by both absorption \citep{qu2022cosmic,chen2023cosmic} and emission lines \citep{chen2023empirical,chen2024ensemble}. However, the complete picture of energy cascade and dissipation, in particular, the origin and impact of turbulence, requires further investigation.

On the theoretical side, early simulations with nonprecessing jets \citep{2006Croton,2007Cattaneo,2011Gaspari, 2017Weinberger} resulted in inefficient energy dissipation. This is why some additional mechanisms such as the ``dentist drill'' \citep{2006Vernaleo} effect have been proposed in the literature to overcome the problem. Subsequent research investigated various jet configurations, including precessing jets \citep{2012Gaspari,2014Lia,2014Lib,2016Yangb,2017Bourne, 2023Husko}, wide-angle jets \citep{2015Prasad,2017Hillel,2018Hillel,2007Sijacki,2008Guo,2012Dubois}, fixed-power jets \citep{2002Bruggen,2004Omma}, and spherical energy injection \citep{2015Reynolds}. However, the values of jet parameters, such as mass flux and opening angle, adopted in the literature are very diverse. Most importantly, these parameters are usually treated as free parameters that are not necessarily consistent with the results obtained by small-scale GRMHD simulations of black hole accretion and jet formation and observational constraints. 

For black hole hot accretion flows, intensive theoretical studies \citep[e.g.,][]{1999MNRAS.303L...1B,2012MNRAS.420.2912B,2015ApJ...799...71G} and numerical simulations \citep{2012ApJ...761..129Y,2012ApJ...761..130Y,   2012Narayan,2015Yuan,2021Yang} have shown that wind is in fact more common than jets. While jets only exist in hot accretion flows around a spinning black hole, wind exists in hot accretion flows around any black hole \citep{2014ARA&A..52..529Y}. The existence of wind and the coexistence of wind and jets have been confirmed by more and more observations \citep[e.g.,][]{2013Sci...341..981W,2014Tombesi,2016Natur.533..504C,2019Park,2021NatAs...5..928S,2022ApJ...926..209S,2024Park}. GRMHD numerical simulations have compared the properties of jets and wind and found that even in the case of a very rapidly spinning black hole, where jets are supposed to be the strongest, the momentum flux of jets is smaller than that of wind, while the total energy flux of jets is larger than that of wind by at most a factor of 10 \citep{2021Yang}. This suggests that wind likely plays an important feedback role thus cannot be neglected. For this reason, wind in the hot feedback mode has been properly incorporated into the MACER framework \citep{Yuan2018}\footnote{But jets are temporarily neglected in this work.}. \citet{2019ApJ...885...16Y} have specifically investigated the role of wind in the hot mode in AGN feedback based on the MACER framework and found that it is especially important in controlling the AGN luminosity. 

In this study, we continue our study of the evolution of an isolated elliptical galaxy, extending \citet{Yuan2018} by including both jets and wind. We find that jets and wind are coupled and their effects are much more important than the sum of jets and wind alone. The general results will be presented in another paper (Guo et al. in preparation). In this paper, we focus on the production of turbulence by the jet-wind interaction and the energy dissipation mechanism of turbulence.   

This paper is organized as follows. In \S\ref{sec:methods}, we describe the numerical setup and physical models, especially the jet model, in our simulations. The main results are presented and analyzed in \S\ref{sec:results}. Finally, we discuss and summarize our findings in \S\ref{sec:implications} and \S\ref{sec:conclusions}.

\section{Methods}
\label{sec:methods}
The hydrodynamic simulations presented in this paper are based on the MACER framework \citep{Yuan2018} and employ 2D axisymmetric spherical coordinates with $240 \times 64$ grid cells in the ($r$, $\theta$) directions. The radial grid spacing is logarithmically increasing, reaching a resolution of $\sim 0.3\,\mathrm{pc}$ at the inner boundary, which lies within the Bondi radius (~$10\,\mathrm{pc}$; \citealt{yao2021}), allowing us to resolve the Bondi radius in our simulations. The computational domain covers a radial range from $2.5\,\mathrm{pc}$ to $500\,\mathrm{kpc}$. The typical resolution within the primary star-forming region (~1 kpc) is about 10 pc. Outflow boundary conditions are applied at the radial boundaries, while reflective boundary conditions are used at the polar angle boundaries.
In our setup, the injected jet velocity corresponds to a Lorentz factor only slightly larger than unity (\(\Gamma \gtrsim 1\)), indicating that relativistic effects are weak. Therefore, the flow can be accurately treated using nonrelativistic hydrodynamics. 

The code ZEUS-MP \citep{2006Hayes} is used in MACER to solve the following governing equations: 
\begin{align}
  \frac{\partial \rho}{\partial t} + \nabla \cdot \left( \rho \bm{v} \right) &= \alpha_\star \rho_{\star} + \dot{\rho}_\mathrm{II}-\dot{\rho}_{\star}^{+}, \label{eq:mom} \\
  \frac{\partial \bm{p}}{\partial t} + \nabla \cdot \left( \bm{p}\bm{v} \right) &= -\nabla p_{\rm gas} + \rho\bm{g} -\nabla p_{\rm rad} - \dot{\bm{m}}_{\star}^{+} \label{eq:energy} \\
  \frac{\partial E}{\partial t} + \nabla \cdot \left( E \bm{v} \right) &= -p_{\rm gas}\nabla \cdot \bm{v} + H - C + \dot{E}_\mathrm{I} + \dot{E}_\mathrm{II} + \dot{E}_{S} - \dot{E}_{\star}^{+},
\end{align}
where $\rho$, $\bm{v}$, $\bm{p}$, and $E$ are the gas density, velocity, momentum, and internal energy, respectively. The gas pressure is determined by the ideal gas equation of state $p_{\rm gas} = \left( \gamma - 1 \right) E$, where the specific heat ratio is $\gamma = 5/3$, and $\bm{g}$ is the gravitational field of the galaxy (i.e., stars, dark matter, plus the contribution of the  central SMBH). Source terms are included in the equations: $\alpha_{\star} \rho_{\star}$ is the mass source from stellar evolution, $\dot{\rho}_\mathrm{II}$ is the recycled gas from supernovae (SNe II), and $\dot{E}_\mathrm{I}$ and $\dot{E}_{II}$ are feedback from SNe Ia and SNe II, respectively \citep{1989Mathews,2005Tang,2012Ciotti,2012Novak}.  The $H$ and $C$ terms denote heating and radiative cooling rates, respectively. The heating/cooling rate calculation follows \citet{2005MNRAS.358..168S} and \citet{2017Xie}, who describe the net heating and cooling per unit volume of the gas in photoionization equilibrium, including Compton heating/cooling, bremsstrahlung cooling, photoionization, and line and recombination cooling. The cooling function assumes solar metallicity, with a cooling temperature floor of $1 \times 10^4\,\mathrm{K}$. \citet{2005MNRAS.358..168S} and \citet{2017Xie} deal with the AGN radiative heating via the calculation of a ``Compton temperature'' in the case of luminous and low-luminosity AGNs, respectively. The $\dot{\bm{m}}_{\star}^{+}$ and $\dot{E}_{\star}^{+}$ terms represent the mass and energy feedback from stellar evolution, respectively, where the age of the stellar population is $2\,\mathrm{Gyr}$, and the star formation criteria and rates are calculated according to previous studies \citep{2011Novak, 1998ApJ...498..541K,2007ApJ...654..304K} (see \citet{2019ARA&A..57..227K} for a review).

The inner boundary of our simulation domain is smaller than the Bondi radius of the black hole accretion. Using the mass flux calculated at the inner boundary combined with black hole accretion theory, we can determine the exact mass accretion rate at the black hole horizon, which is the most important parameter for AGN feedback studies since it determines the AGN power. Moreover, the state-of-the-art AGN physics has been incorporated, again thanks to the small inner boundary. Our simulations study the evolution of an isolated elliptical galaxy with stellar mass $M_\star = 3 \ \times 10^{11} M_\odot$ and central supermassive black hole (SMBH) mass $M_\mathrm{BH} = 4.5\ \times 10^9 M_\odot$. The galaxy model has a spherically symmetric gravitational potential from stars (Jaffe profile) and dark matter. \

There are two AGN accretion/feedback modes, with hot and cold modes separated by a critical luminosity $L_{c} \sim 2\% L_\mathrm{Edd}$ \citep{Yuan2018}. The corresponding critical mass accretion rate is $\mdot_{c} \approx L_{c}/(\epsilon_\mathrm{EM,cold} c^{2})$ where $\epsilon_\mathrm{EM,cold}$ is the radiative efficiency. The wind exists in both cold and hot modes, characterized by wide opening angle and low speed compared to jets. The parameters of cold wind are taken from observational statistical results, while the parameters of hot wind are taken from small-scale GRMHD simulations of black hole accretion by \citet{2015Yuan}.

In hot mode, we consider the most updated mass flux angular distribution and half-opening angle of the hot wind which calculated by \cite{2021Yang} to study realistic AGN feedback physics. The half opening-angle of hot wind is $50^{\circ}$ in hot mode.

The fluxes of mass, momentum, and energy of hot wind on
the scale black hole accretion were carefully studied based on 3D GRMHD simulation \citep{2015Yuan},
\begin{equation}
    \dot{M}_{\rm W,H} \approx \dot{M}{(r_{\rm in})} \left[ 1 - \left( \frac{3\,r_{s}}{r_{\rm tr}} \right)^{0.5} \right],
\end{equation}
where $\dot{M}{(r_{\rm in})}$ is the mass accretion rate at the innermost radius of the simulation domain and the Schwarzschild radius is $r_s \equiv 2G M_{\mathrm{BH}}/c^2$. The $r_{\rm tr}$ is the radius where the thin disk will be truncated and transit into a hot accretion flow. The value of the transition radius is \citep{2014Yuan,Yuan2018}:
\begin{equation}
r_{\rm tr}=3r_{\rm s}\left[ \frac{2\times10^{-2}\dot M_{Edd}}{\dot M(r_{\mathrm{in}})}\right]^2
\end{equation}
where the $\dot M_{Edd}$ is the Eddington accretion rate.
\begin{equation}
    v_{\rm W,H}\approx 0.64\,v_{\rm K}(r_{tr}),
\end{equation}
where $v_{\rm K}(r_{tr})$ is the Keplerian velocity at transition radius.

In cold mode, the properties of wind to be used in the present work are derived from observations. The half-opening angle of the cold wind is $\sim45^{\circ}$, the same as in previous MACER models \citep{2011Novak,2017Ciotti,Yuan2018}. In this paper, we adopt the fitted formulas for mass fluxes and velocities of cold winds presented in \citet{2015Gofford} to describe the wind:
\begin{align}
      \dot{M}_{\rm W,C} &= 0.28 \left(\frac{L_{\mathrm{BH}}}{10^{45} \mathrm{erg\ s^{-1}}} \right)\ \dot M_{\odot}\ \mathrm{yr^{-1}}\\
     v_{\mathrm{W,C}}&=2.5\times 10^4\ \left( \frac{L_{\mathrm{BH}}}{10^{45} \mathrm{erg\ s^{-1}}} \right)\  \mathrm{km\ s^{-1}}
\end{align}
where the $L_{\mathrm{BH}}$ is the bolometric luminosity of the AGN.

The key new component of the feedback in this work is the AGN jet in hot mode, which was neglected in \citet{Yuan2018}. Different from many previous works where jet parameters are treated as free parameters, in our model the jet parameters are  taken from the small-scale GRMHD simulation of black hole accretion and jet formation by \citet{2021Yang}. The detailed setup of the jet model is described in Guo et al. (2025, in preparation) and here we only briefly describe it. The jet is modeled as a relativistic outflowing fluid ejected perpendicular to the accretion flow. The important parameters of the jet include a half-opening angle of $7.5^\circ$, a peak velocity of $0.5 c$, and a mass flux of $\dot M_\mathrm{jet} = 0.35 \dot{M}_\mathrm{BH}$ where the BH accretion rate $\dot{M}_\mathrm{BH}$ is the mass accretion rate at the black hole horizon: $\dot M_{\rm BH}=\dot M_{\rm acc}- \mdot_{\rm jet}-\dot M_\mathrm{wind, hot}$, where $\dot M_{\rm acc}$ is the mass accretion rate at the inner boundary of our simulation domain and $\dot M_\mathrm{wind, hot}$ is the hot wind mass flux. There is no jet precession in our model. The jet is activated only when the AGN is in the hot mode, i.e., when the mass accretion rate is lower than the critical value $\mdot_{c}$.

Our simulation suite comprises three representative runs: {\tt FullFeedback}, {\tt WindOnly}, and {\tt JetOnly}. As their names suggest, the {\tt FullFeedback} run includes both jet and wind feedback, while the others include only wind or jet feedback, respectively. Since jets are present only during the hot accretion mode, jet feedback is activated in {\tt FullFeedback} and {\tt JetOnly} runs exclusively when the AGN enters this mode. It is important to note that in the {\tt JetOnly} run, we disable the wind only during the hot mode (when jets are present) while retaining it during the cold mode (when jets are absent), as wind is always present in any mode.  All simulations are evolved for $t = 12\,\mathrm{Gyr}$.

\section{Results}
\label{sec:results}

\subsection{Spatial distribution of gas properties}

\begin{figure*}
    \centering
    \includegraphics[width=0.97\textwidth]{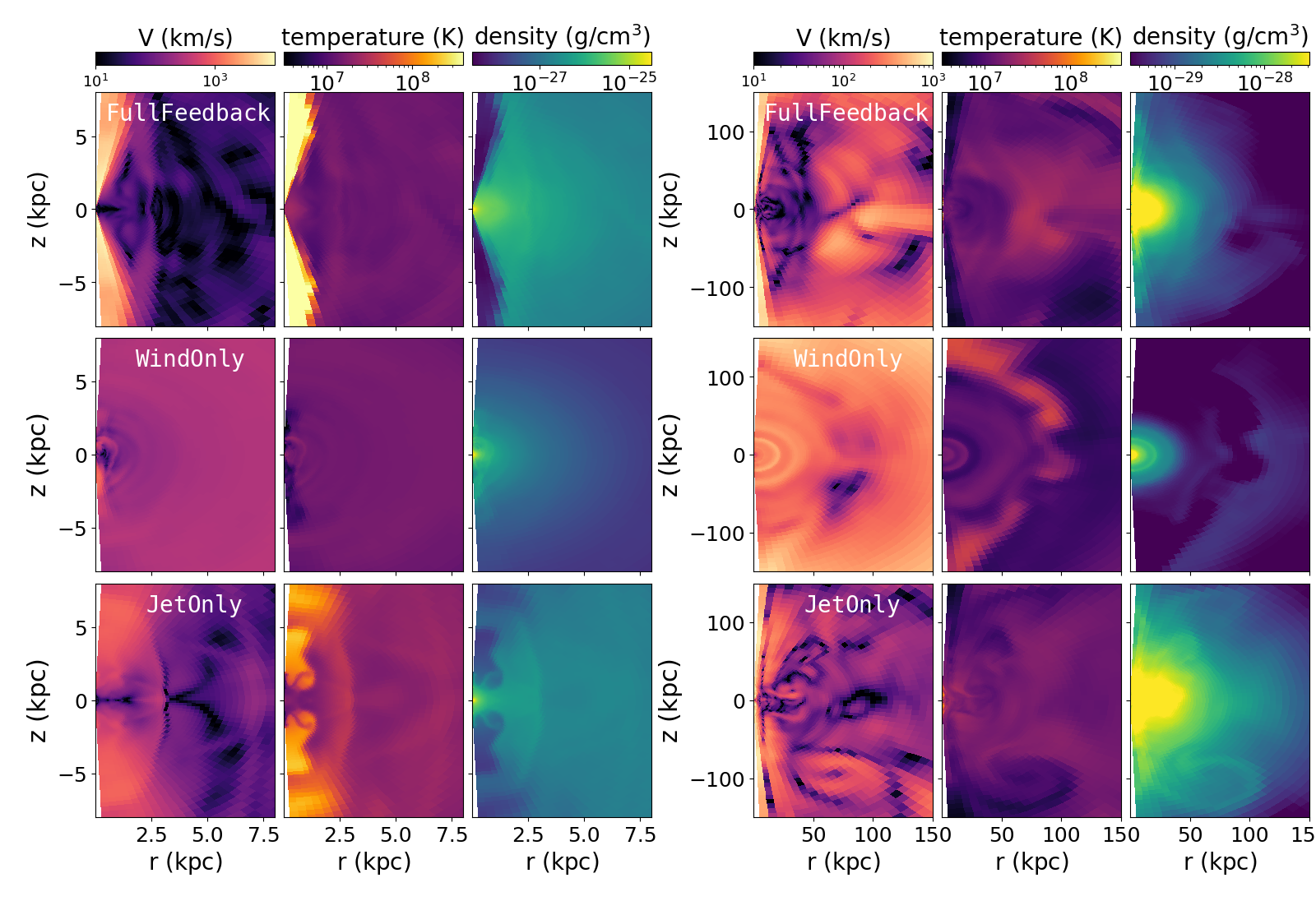}
    \caption{The spatial distribution, at scales of $\sim 8\,\mathrm{kpc}$ (left panel) and $\sim 150\,\mathrm{kpc}$ (right panel), of gas velocity (left), temperature (middle), and density (right) in the simulations {\tt FullFeedback} (top), {\tt WindOnly} (middle) and {\tt JetOnly} (bottom). Compared with {\tt WindOnly}, which is more isotropic, the jet-included runs {\tt FullFeedback} and {\tt JetOnly} show a more anisotropic distribution of gas properties. The jet in the {\tt JetOnly} run is less collimated than in the {\tt FullFeedback} run, where the jet is wrapped by the hot wind.}
  \label{fig:gasproperty}
\end{figure*}
Since the simulations span a wide range of spatial scales, we first present a qualitative view of the spatial distribution of gas properties at two different scales: $\sim 8\,\mathrm{kpc}$ and $\sim 150\,\mathrm{kpc}$, as shown in Fig.~\ref{fig:gasproperty}. The gas velocity magnitude (left), temperature (middle), and density (right) are shown for the {\tt FullFeedback} (top), {\tt WindOnly} (middle), and {\tt JetOnly} (bottom) runs. As expected, the strong anisotropy in gas properties arises in the {\tt FullFeedback} and {\tt JetOnly} runs with jets included. However, the jet morphologies in the two runs are distinct at kpc scales: the jet in {\tt FullFeedback} is well collimated, and the jet regions near the vertical axis are dominated by hot (up to $10^{9}\,\mathrm{K}$), low-density, and high-velocity (up to a few $10^3\,\mathrm{km/s}$) gas. In contrast, the jet in {\tt JetOnly} is more expanded with a broader opening angle, forming a cocoon-like structure and showing shock-wave features. For the {\tt WindOnly} run, the gas properties are more isotropic, with the gas temperature and density showing a more uniform distribution. The gas velocity is also more isotropic, with a lower magnitude compared to the jet-included runs. At the larger scale of $\sim 150\,\mathrm{kpc}$, the anisotropy in gas properties is still evident in the jet-included runs, with the jet in {\tt FullFeedback} showing a more collimated structure compared to the jet in {\tt JetOnly}. The hot wind in the {\tt FullFeedback} run is also more isotropic, with a spatially narrower distribution of gas properties compared to the jet-included runs.

\begin{figure}
    \centering
    \includegraphics[width=0.5\textwidth]{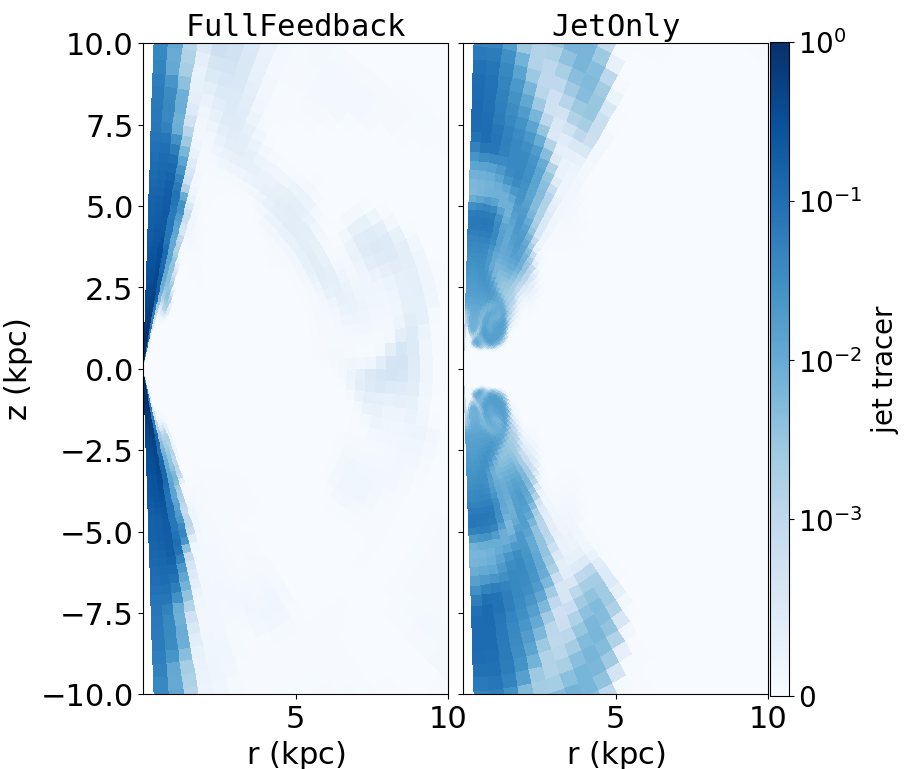}
    \caption{ The spatial distributions of AGN jet tracer within 10 kpc in the {\tt FullFeedback} (left) and {\tt JetOnly} (right) simulations. The jet is well-collimated in {\tt FullFeedback} by the surrounding AGN hot wind, while the jet in {\tt JetOnly} is more expanded and less collimated, showing intermittent structures due to shocks.}
    \label{fig:jetcolli}
\end{figure}

Fig.~\ref{fig:jetcolli} shows the spatial distribution of the AGN jet tracer within $10\,\mathrm{kpc}$ in the {\tt FullFeedback} (left) and {\tt JetOnly} (right) simulations. The jet tracer is a passive scalar that traces the jet material injected at the inner boundary. The jet in the {\tt FullFeedback} run is well-collimated by the surrounding AGN hot wind, showing a more confined and narrow structure. This result is consistent with \citet{2019Park}. The supersonic jets are dominated by a ram pressure $P_\mathrm{ram}$ of a few $10^{-1}\,\mathrm{eV\,cm^{-3}}$, comparable to the total pressure of the surrounding hot wind. Beyond $\sim 7\,\mathrm{kpc}$, the outskirts of the jet tracer start to smear out, indicating the mixing of the jet with the hot wind due to turbulent entrainment. In contrast, the jet in the {\tt JetOnly} run is more expanded and less collimated, showing intermittent structures due to shocks.

As a quick recap, for jet versus non-jet runs, the inclusion of the AGN jet feedback can change the thermal and kinematic properties of the gas, which is not surprising; however, a more interesting hint is that the effects of the jet feedback seem to strongly correlate with the environment in which the jet lives, i.e., whether the jet is wrapped by the hot wind or not. This is already evidenced by the collimation of the jet, as shown in Fig.~\ref{fig:jetcolli}, and its implications will be further discussed in depth in the following sections.

\subsection{Gas phases in central regions}

\begin{figure*}
    \centering
    \includegraphics[width=0.95\textwidth]{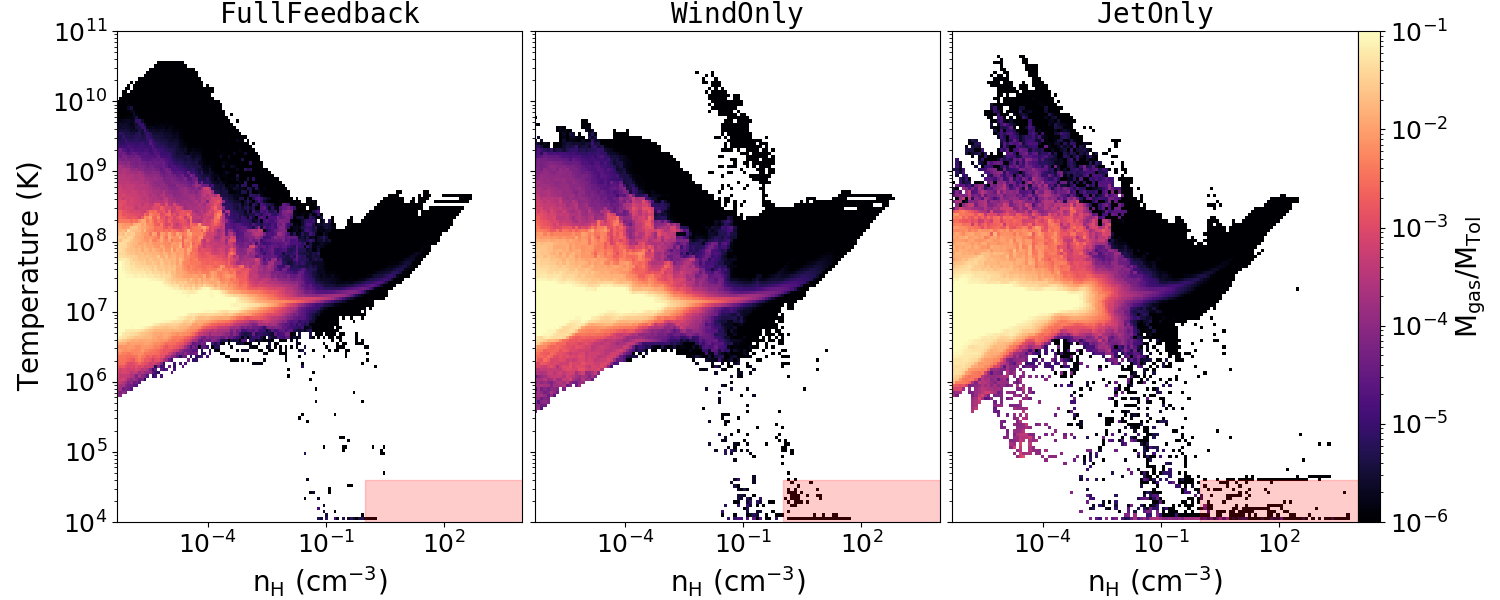}
    \caption{The density vs. temperature phase distributions of the gas within $35\,\mathrm{kpc}$ from different simulations: {\tt FullFeedback} (left), {\tt WindOnly} (middle), and {\tt JetOnly} (right), where the color is weighted by the gas mass and averaged over the entire simulation time ($12\,\mathrm{Gyr}$). The red shaded areas in the lower-right corners denote the density-temperature space where the star formation threshold is satisfied. The jet-included runs ({\tt FullFeedback} and {\tt JetOnly}) are more effective in lifting gas to higher temperatures, and the {\tt FullFeedback} run with both wind and jets included dramatically reduces the mass of star-forming cool gas compared to the other two runs.}
    \label{fig:gasphase}
\end{figure*}

To investigate the impact of AGN feedback, in particular the jet feedback, on the gas phases and star formation of the galaxy, in Fig.~\ref{fig:gasphase} we examine the gas density vs. temperature phase distributions within $r \leq 35\,\mathrm{kpc}$ from different simulations: {\tt FullFeedback} (left), {\tt WindOnly} (middle), and {\tt JetOnly} (right), where the phase distributions are volume-weighted and averaged throughout the entire simulation time ($12\,\mathrm{Gyr}$). Superposed on the phase plots are red shaded areas in the lower-right corners of the density-temperature space where the star formation threshold is satisfied, i.e., $T \leq 4\times10^{4}$ K and $n_\mathrm{H} \geq 1\,\mathrm{cm^{-3}}$.

A few key features are observed in the phase plots. First, in all three simulations, the gas distributions in the density-temperature space significantly expand by orders of magnitude around the initial hydrodynamic equilibrium state, indicating that the gas reaches some degree of quasi-stable state but is significantly perturbed by feedback processes as indicated by the large scattering in the phase space. In particular, compared with the {\tt WindOnly} run without jets, gas in the jet-included runs ({\tt FullFeedback} and {\tt JetOnly}) is more effectively heated and uplifted to higher temperatures of $\gtrsim 10^{9.5}\,\mathrm{K}$, as evidenced by the high-temperature, relatively lower-density gas in the upper-left region of the phase plot. This suggests that the jet feedback is more effective in heating the gas due to its much higher energy injection rate than the wind feedback.

Second, focusing on the red-shaded regions in the lower-right corner, we observe that the gas distribution within the red-shaded star-forming region in {\tt JetOnly} is significantly enhanced compared to that in {\tt WindOnly} and {\tt FullFeedback}, with $\sim 10^9\ M_\odot$ of newly formed stars in the {\tt JetOnly} run. This indicates that the jet feedback alone is ineffective in reducing star formation, which is expected due to the narrow opening angle and localized heating of the jet. The {\tt WindOnly} simulation, on the other hand, shows a slightly lower mass of newly formed stars, $\sim 2 \times 10^8\ M_\odot$, compared to the {\tt JetOnly} run, since the hot wind has a larger opening angle and can isotropically heat the gas. In contrast, the {\tt FullFeedback} simulation exhibits a significantly reduced mass of newly formed stars, $\sim 9 \times 10^6\ M_\odot$, which is $1$ -- $2$ orders of magnitude lower than the other two simulations, indicating that neither jets nor wind feedback alone can effectively suppress star formation, but the combination of both can.

We emphasize that the inefficiency of star formation suppression in {\tt JetOnly} is not due to reduced AGN energy output; on the contrary, the total AGN energy output across the three simulations reveals {\tt JetOnly} $\sim 1 \times 10^{62}$ erg, {\tt FullFeedback} $\sim 4 \times 10^{61}$ erg, and {\tt WindOnly} $\sim 3 \times 10^{60}$ erg. This demonstrates that although the {\tt JetOnly} simulation injects 1--2 orders of magnitude more energy than the other two simulations, it remains the least effective in suppressing star formation. This implies that high total energy output does not necessarily lead to efficient star formation suppression; rather, energy must be deposited in a dissipative manner into the ISM to be effective.

\begin{figure*}
    \centering
    \includegraphics[width=0.95\textwidth]{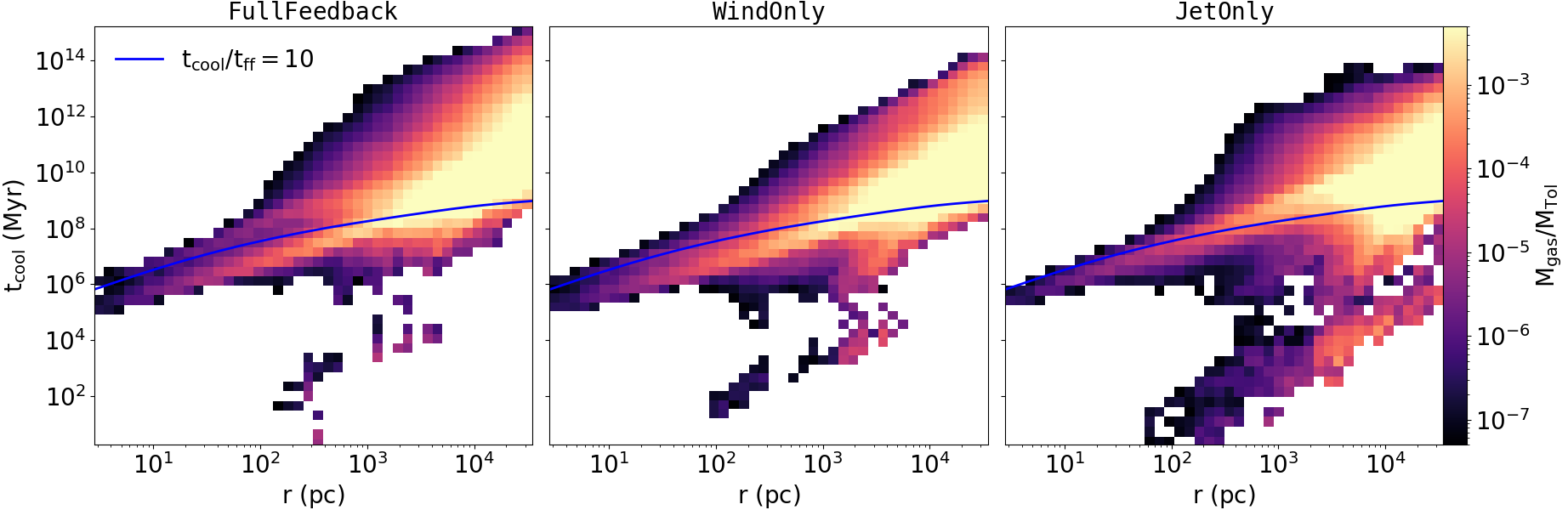}
    \caption{The mass-weighted distribution of gas cooling time $t_\mathrm{cool}$ vs. radius $r$ in the {\tt FullFeedback}, {\tt WindOnly} and {\tt JetOnly} runs, superposed by blue lines representing $t_\mathrm{cool}/t_\mathrm{ff}=10$. The {\tt JetOnly} run shows significantly more gas with $t_\mathrm{cool}/t_\mathrm{ff}<10$ than the other two runs, indicating that the jet feedback alone is less effective in heating the gas and suppressing the condensation process.}
    \label{fig:tc/tff}
\end{figure*}

The origin of the star-forming cool gas is closely related to the gas cooling time. Fig.~\ref{fig:tc/tff} shows the mass-weighted gas cooling time ($t_\mathrm{cool}$) versus radius $r$ in the {\tt FullFeedback}, {\tt WindOnly}, and {\tt JetOnly} runs, superposed by blue lines representing $t_\mathrm{cool}/t_\mathrm{ff}=10$, where $t_\mathrm{ff}$ is the free-fall time. The gas below the blue line with $t_\mathrm{cool}/t_\mathrm{ff} \lesssim 10$ is susceptible to thermal instability and can condense to form cold clumps \citep{2012Sharma,2015Meece,2017Voit}. The {\tt JetOnly} run shows significantly more gas with $t_\mathrm{cool}/t_\mathrm{ff}<10$ than the other two runs, indicating that the jet feedback alone is less effective in heating the gas and suppressing the condensation process. We note that while shock-induced compression could potentially enhance cooling and trigger star formation in denser environments, this effect is not observed at the vicinity of shocks in our elliptical galaxy setup, due to the relatively low gas densities and long cooling timescales, which prevent shock compression from substantially enhancing cooling rates and star formation. In contrast, our recent study using the latest 3D MACER framework to investigate AGN feedback in starburst dwarf galaxies \citep{su2025positive} demonstrates that shock-enhanced cooling can indeed lead to enhanced star formation, and thus positive AGN feedback when gas densities are sufficiently high, cooling times are short, and the AGN is not powerful enough to efficiently clear the ISM.

\subsection{Gas entropy and density profiles}

\begin{figure*}
    \centering
   \includegraphics[width=0.75\textwidth]{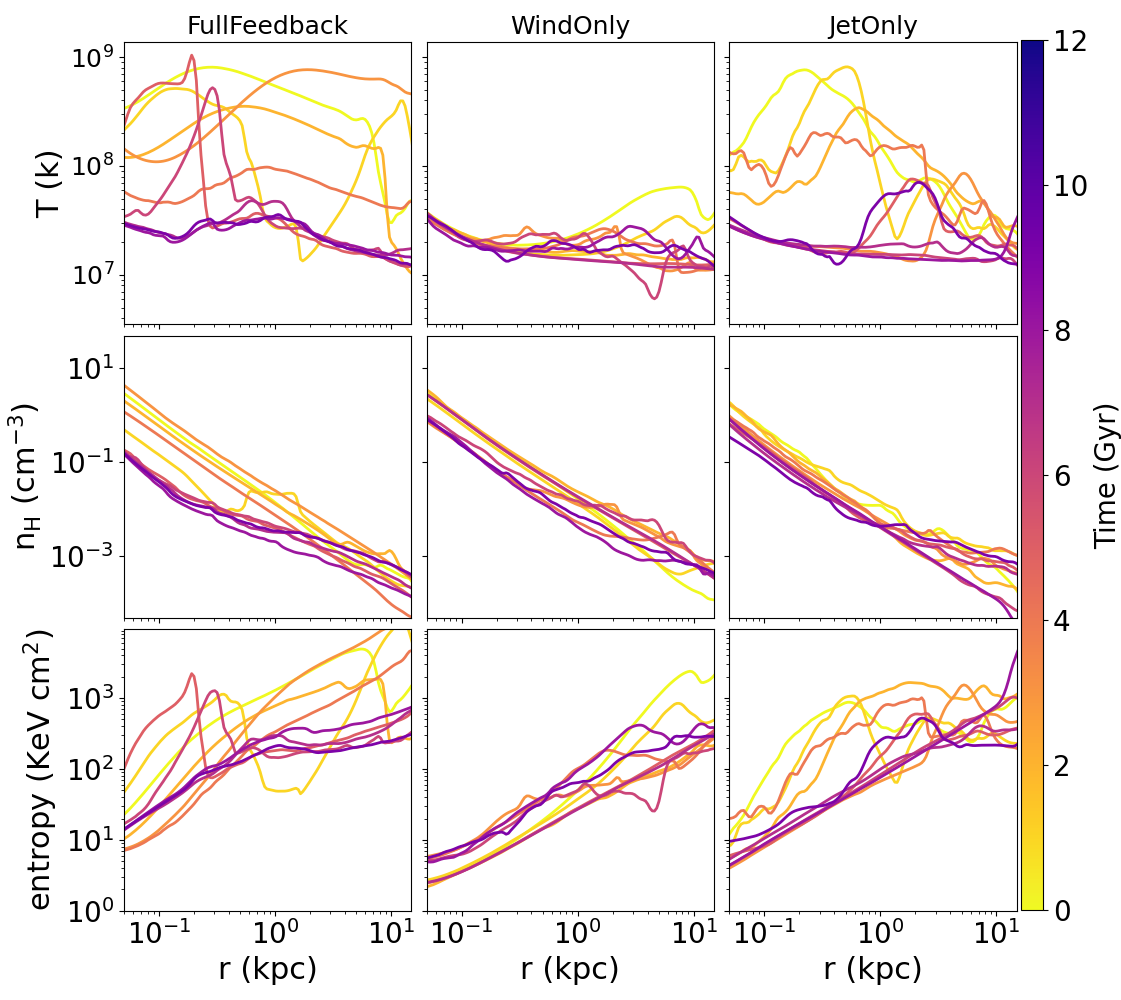}
    \caption{Time evolution of gas temperature (top), number density (middle) and entropy (bottom) profiles in {\tt FullFeedback} (left), {\tt WindOnly} (middle) and {\tt JetOnly} (right) runs. Colors represent different times, ranging from 0 to 12 Gyr, as indicated by the color bar.\label{fig:K profiles}}
\end{figure*}

Fig.~\ref{fig:K profiles} presents the profiles of gas temperature, number density and specific entropy at different times in the {\tt FullFeedback} (left), {\tt WindOnly} (middle), and {\tt JetOnly} (right) runs.
In all simulations, the profiles of gas entropy and number density deviate from their initial states and reach a quasi-steady state after $\sim 6\,\mathrm{Gyr}$, indicating that the AGN feedback has effectively regulated the gas properties in the galaxy.

Differences in the entropy and density profiles among the three simulations are evident. In the {\tt FullFeedback} run, the gas entropy is higher than the other two simulations and the gas density is lower, with the density profile flattening out at smaller radii, indicating that the AGN feedback is effective in heating the gas and suppressing star formation. In contrast, the {\tt WindOnly} run shows a less pronounced increase in entropy from its initial state. The density profiles in the {\tt WindOnly} and {\tt JetOnly} runs show a power-law-like distribution at smaller radii.

Compared to the X-ray observations of galaxy clusters \citep{2009Cavagnolo}, galaxy groups and giant ellipticals \citep{2012Werner,2014Werner}, all three models show values are similar to the observed non cool-core entropy profiles in \citet{2009Cavagnolo}, and the simulations with jets ({\tt FullFeedback} and {\tt JetOnly}) show higher entropy than {\tt WindOnly}. One reason is we assume a black hole spin of $a = 0.98$, allows the black hole to drive powerful outflows, since the magnetic field lines in the jet region are anchored to the black hole ergo-sphere and can extract the spin energy via the Blandford-Znajek process \citep{1977Blandford}, with $P_{\mathrm{jet}} \propto a^2$ \citep{2019Chatterjee}, in the magnetically arrested disks (MAD) state \citep{2021Yang}. We make this choice to study how a more powerful AGN in elliptical galaxies affects galactic evolution. Another reason is that we do not consider environmental effects, focusing primarily on isolated elliptical galaxies, which may lead to the absence of cosmological inflow. Consequently, the average gas density is lower and the cooling efficiency is also reduced. These two factors work together to produce higher entropy profiles in {\tt FullFeedback} and {\tt JetOnly}.

\subsection{Star formation}

The star formation model in this work is calculated by subtracting the gas that satisfied the star formation threshold ($n_{\mathrm{H}}\geq 1\ \mathrm{cm^{-3}}$ and $T\leq 4\times10^4\ K$) as a star formation rate per unit volume\citep[see][for more details]{2011Novak}:
\begin{equation}
    \dot{\rho}_{\rm SF} = \frac{\eta_{\rm SF}\,\rho}{\tau_{\rm SF}},
\end{equation}
where we adopt the star formation efficiency of $\eta_{\rm SF}=0.1$, and the star formation time scale,
$\tau_{\rm SF}$ is:
\begin{equation}
    \tau_{\rm SF} = \max(\tau_{\rm cool},\tau_{\rm dyn}),
\end{equation}
where the cooling time scale, $\tau_{\rm cool}$, and the dynamical time scale, $\tau_{\rm dyn}$ are:
\begin{equation}
    \tau_{\rm cool} = \frac{E}{C},~~ \tau_{\rm dyn}=\min(\tau_{\rm ff},\tau_{\rm rot}),
\end{equation}
with
\begin{equation}
    \tau_{\rm ff} = \sqrt{\frac{3\,\pi}{32G\rho}},~~ \tau_{\rm rot}=\sqrt{r\frac{\partial\,\Phi(r)}{\partial\,r}},
\end{equation}
where $E$ is the internal energy density, $C$ is the cooling rate per unit volume, and $\Phi(r)$
is the gravitational potential at a given radius.

\begin{figure}
    \centering
    \includegraphics[width=0.95\linewidth]{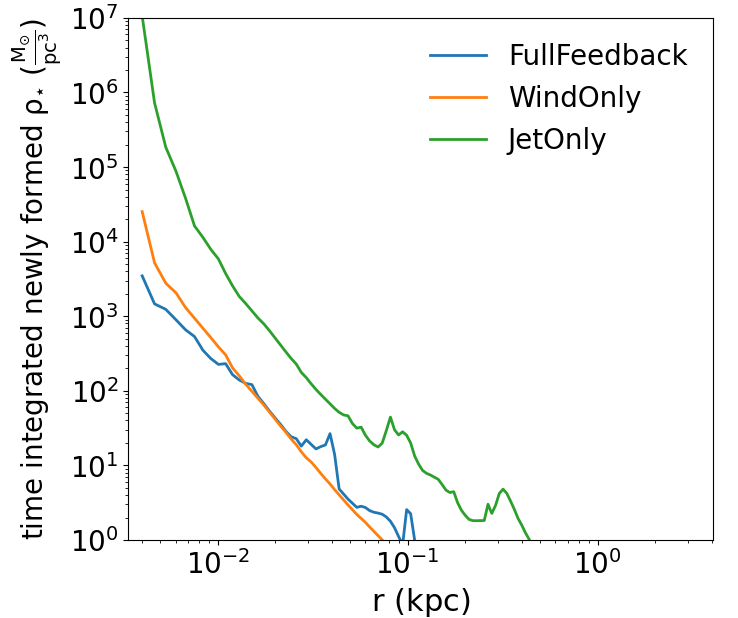}

    \caption{Radial distribution of the mass density of newly formed stars during 0-12 Gyr. The green lines show the star formation of {\tt JetOnly}, blue lines for {\tt FullFeedback} and yellow lines for {\tt WindOnly} simulations. The {\tt JetOnly} simulation shows a much higher star formation rate and more extended star formation region.}
    \label{fig:interSFR}
\end{figure}

Fig.~\ref{fig:interSFR} shows the radial profiles of new stellar mass density, where the mass, integrated over the entire simulation time of $12\,\mathrm{Gyr}$, is plotted against the spherical radius $r$ up to $3\,\mathrm{kpc}$. Among the three simulations, {\tt JetOnly} is the most ineffective in suppressing star formation with the highest total mass of newly formed stars of $10^8\,M_\odot$, barely remaining quenched. The {\tt WindOnly} simulation shows a slightly lower mass of newly formed stars of $10^7\,M_\odot$, with a smooth radial density profile of newly formed stars, suggesting a lack of compressed gas by strong shocks. The results of {\tt JetOnly} and {\tt WindOnly} simulations suggest that neither jet feedback nor wind feedback alone is able to suppress star formation effectively. For comparison, a pure cooling simulation without any feedback \citep{Yuan2018} produces a total mass of newly formed stars of $10^{10}\,M_\odot$, demonstrating the critical importance of AGN feedback in regulating star formation.

In contrast to the other two simulations, the {\tt FullFeedback} simulation exhibits significantly distinct behaviors in two aspects. First, the total mass of newly formed stars is dramatically reduced below $10^{6}\,M_\odot$, which is $1$ -- $2$ orders of magnitude lower than the other two simulations. Second, the radial profile of new stellar mass density is much lower than the other two simulations within a radius of $r\sim 0.1\,\mathrm{kpc}$, indicating that star formation in the central region is effectively suppressed by the combined wind and jet feedback. The results of star formation are consistent with the gas entropy and density profiles, where the {\tt FullFeedback} simulation shows the most effective heating and suppression of gas condensation, leading to the lowest star formation rate and the most concentrated star-forming region. Furthermore, the {\tt FullFeedback} simulation demonstrates that the synergistic effects of jets and winds are more effective in suppressing star formation than either feedback mechanism alone. We will further discuss this synergy quantitatively in the following sections.

\subsection{Jet-wind shears and the resulted turbulence}

\begin{figure*}
    \centering
    \includegraphics[width=\columnwidth,height=0.4\textheight]{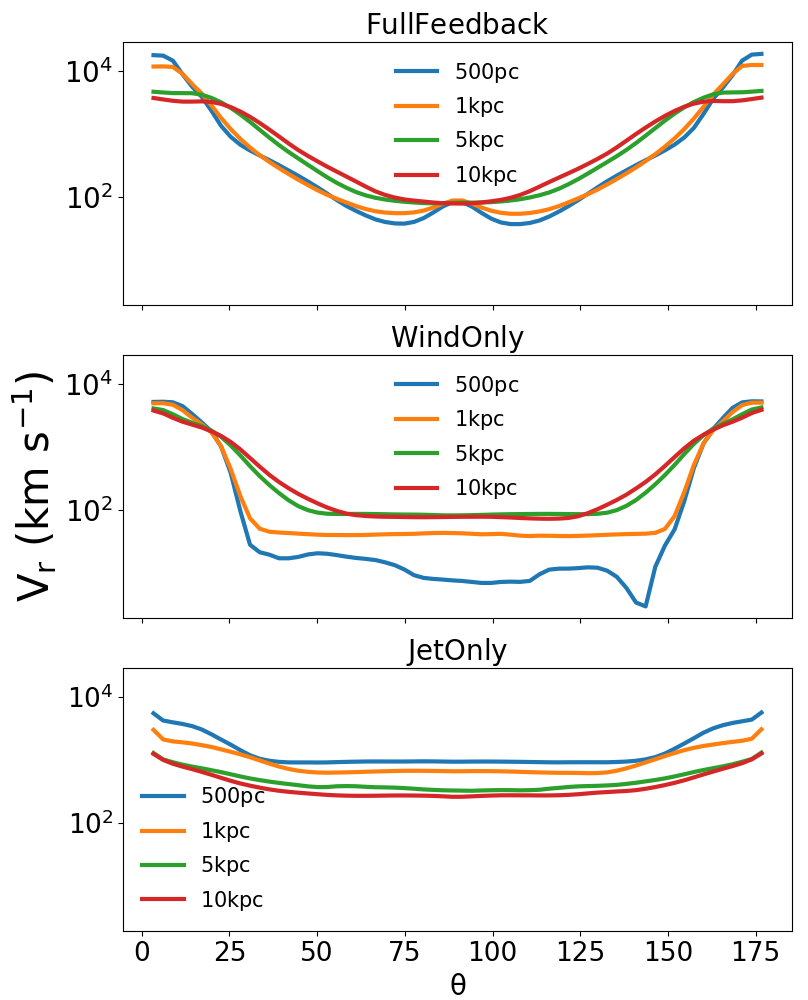}
    \hfill
    \includegraphics[width=\columnwidth,height=0.4\textheight]{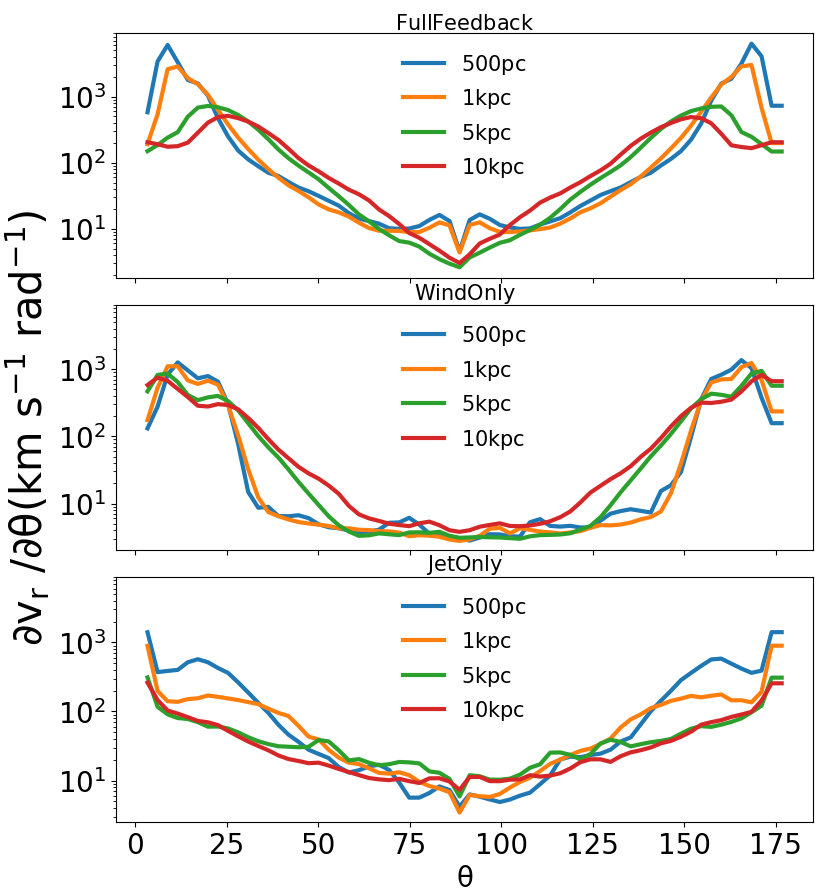}
    \caption{Radial velocity ($v_r$) profiles as a function of polar angle ($\theta$)(left) and velocity shear ($\partial v_r/\partial \theta$) profiles as a function of polar angle ($\theta$)(right) for the {\tt FullFeedback} (top), {\tt WindOnly} (middle), and {\tt JetOnly} (bottom) simulations, time-averaged over the entire $12\,\mathrm{Gyr}$ simulation period. Different colors represent velocity profiles at various radii. The {\tt FullFeedback} simulation exhibits the strongest velocity shear which drives efficient KH instabilities and turbulence generation, while the {\tt JetOnly} simulation shows substantially flatter velocity profiles.}
    \label{fig:shear}
\end{figure*}
Previous sections have demonstrated that the {\tt FullFeedback} simulation significantly outperforms the other simulations in suppressing cool gas formation, particularly within the radial range of $1$--$10\,\mathrm{kpc}$ (Fig.~\ref{fig:tc/tff}), thereby inhibiting star formation and maintaining the galaxy in a quenched state. The amount of cold gas ultimately depends on the competition between heating and cooling in the ISM. A larger ratio of cooling to heating rate will result in more cold gas, which in turn significantly enhances the heating by AGN feedback (since it enhances the mass accretion rate of AGN). For such a ``self-regulation'' system, the efficiency of converting AGN power to the heating rate should be a crucial parameter for determining the amount of cold gas. We find that shear between jets, wind, and ISM produces turbulence through the Kelvin-Helmholtz (KH) instability, and we assume that turbulence dissipation is the main mechanism of converting AGN power to the heating of the ISM. In the following, we compare the efficiency of the three models.  

Figure~\ref{fig:shear} presents the radial velocity ($v_r$) profiles as a function of polar angle ($\theta$) for the {\tt FullFeedback} (top), {\tt WindOnly} (middle), and {\tt JetOnly} (bottom) simulations, time-averaged over the entire $12\,\mathrm{Gyr}$ simulation period. At each snapshot, different colors represent the radial velocity $v_r$ at various radii ranging from $0.5\,\mathrm{kpc}$ to $10\,\mathrm{kpc}$. The {\tt FullFeedback} simulation exhibits pronounced velocity gradients across all sampled radii, indicating persistent strong shear flows throughout the entire angular domain. In contrast, the {\tt WindOnly} simulation displays more moderate shear profiles with diminishing gradients at larger radii. The {\tt JetOnly} simulation, however, demonstrates even flatter velocity profiles due to poor jet collimation, suggesting that isolated jet feedback is considerably less effective in generating the shear flows necessary for turbulence development.

Based on these velocity distributions, we can approximate the growth rate of the KH instability using the relation $\gamma \sim \chi^{1/2} k v_\mathrm{shear}$, where $\chi$ represents the density contrast between adjacent shear layers, $k$ denotes the perturbation wavenumber, and $v_\mathrm{shear}$ corresponds to the velocity differential across the shear interface. While the density contrast $\chi$ across the angular direction remains comparable among all three simulations (with the exception of localized low-density regions within the narrow jet opening angle in {\tt FullFeedback}), the shear velocity $v_\mathrm{shear}$ differs significantly. In the {\tt FullFeedback} simulation, $v_\mathrm{shear} \sim v_r$, substantially exceeding that of the {\tt JetOnly} simulation where $v_\mathrm{shear} \ll v_r$. This disparity translates into fundamentally different timescale relationships: $t_\mathrm{KH} \sim t_\mathrm{dynamic}$ in {\tt FullFeedback} versus $t_\mathrm{KH} \gg t_\mathrm{dynamic}$ in {\tt JetOnly}, where $t_\mathrm{dynamic} \sim r/v_r$ represents the characteristic timescale for the propagation of the outflow. Consequently, the shear flows in the {\tt FullFeedback} simulation generate turbulence more effectively, while turbulent dissipation ultimately leads to more efficient heating and suppression of cool gas formation.

\subsection{Turbulent properties: solenoidal vs. compressive modes}

To further quantify the properties of turbulence driven by different feedback channels, we investigate the solenoidal and compressive modes of the velocity field. The solenoidal mode, a divergence-free term calculated as $\nabla \times \bm{v}$, is associated with rotational motions and responsible for generating turbulence, while the compressive mode $\nabla \cdot \bm{v}$ is associated with converging and diverging motions and responsible for shock generation. Therefore, the ratio of solenoidal to compressive modes of the velocity field is a useful diagnostic to understand the efficiency of the feedback in generating shocks and turbulence. \footnote{We note that in our axisymmetric 2D simulations, equipartition between solenoidal and compressive modes corresponds to a 1:1 ratio, since the azimuthal component does not contribute an additional degree of freedom for velocity fluctuations. This differs from 3D simulations where equipartition yields a 2:1 ratio.}

\begin{figure}
    \centering
    \includegraphics[width=0.45\textwidth]{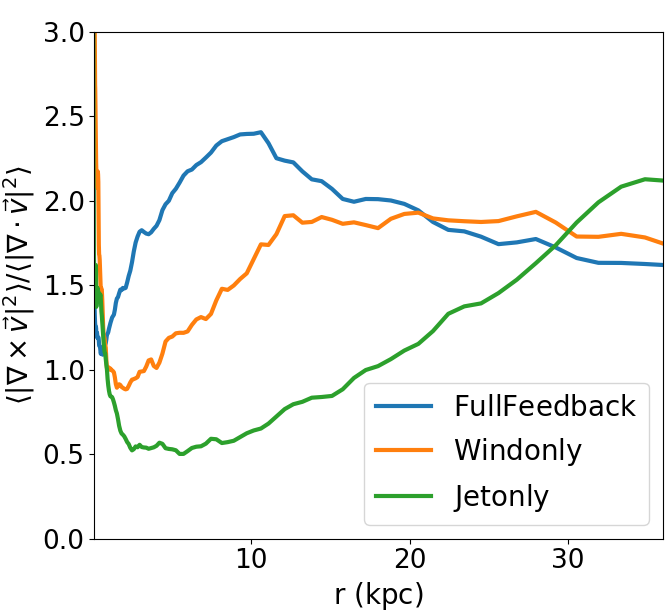}
    \caption{The ratio of solenoidal to compressive modes of the velocity field versus the radius in the {\tt FullFeedback} (blue), {\tt WindOnly} (orange), and {\tt JetOnly} (green) simulations, averaged over the entire simulation time of $12\,\mathrm{Gyr}$. The solenoidal mode dominates the velocity field in both {\tt FullFeedback} and {\tt WindOnly} simulations, while the compressive mode dominates the velocity field in the {\tt JetOnly} simulation within $\sim 20\,\mathrm{kpc}$.}
    \label{fig:solcomp}
\end{figure}
Fig.~\ref{fig:solcomp} shows the ratio of solenoidal to compressive modes of the velocity field, i.e., $\left\langle |\nabla \times \bm{v}|^2 \right\rangle / \left\langle |\nabla \cdot \bm{v}|^2 \right\rangle$, versus the radius in the {\tt FullFeedback}, {\tt WindOnly}, and {\tt JetOnly} simulations. The velocity fields in both {\tt FullFeedback} and {\tt WindOnly} are dominated by solenoidal modes for all the radii shown in the figure, indicating that the turbulence is prevalent in these simulations. Within the central region of $\lesssim 20\,\mathrm{kpc}$, {\tt FullFeedback} shows a slightly higher ratio of solenoidal to compressive modes compared to {\tt WindOnly}, even though the jets are present in {\tt FullFeedback}, which is expected to generate more compressive modes. This suggests that the winds in the {\tt FullFeedback} simulation are more effective in generating turbulence than the jets. The {\tt JetOnly} simulation, on the other hand, shows a different behavior: the compressive mode dominates the velocity field within $\sim 20\,\mathrm{kpc}$, suggesting that the jet feedback alone is more effective in generating shocks than turbulence. Beyond $\sim 20\,\mathrm{kpc}$, the solenoidal mode starts to dominate the velocity field in {\tt JetOnly}, and the compressive modes gradually dissipate via shock thermalization. This is consistent with the results discussed in the previous sections: in {\tt FullFeedback}, winds confine the jet and maintain collimation over larger scales, generating sustained velocity shears (Fig.~\ref{fig:shear}) with KH growth timescales comparable to local dynamical timescales, providing optimal conditions for solenoidal mode generation. In contrast, the {\tt JetOnly} jet rapidly expands into low-density regions, depositing energy primarily through shock thermalization and producing predominantly compressive modes.

\subsection{Turbulent properties: power spectrum and energy dissipation}

To analyze the turbulent properties arising from AGN jet-wind interaction, we examine the power spectrum of turbulent energy and velocity. Due to the logarithmic spacing of grid points in the radial direction, calculating the power spectrum requires careful consideration. Direct interpolation to a uniform grid would introduce numerical artifacts, necessitating an alternative approach. Rather than summing the power spectra across radial regions, we analyze each radial shell individually, accounting for aliasing effects before plotting the spectra versus their corresponding wavenumbers and turbulent kinetic energy. This approach is appropriate because the wavenumber ranges of different shells partially overlap, preventing artificial amplification or suppression.
We note that due to the limited angular resolution, the inertial range of the turbulence cascade is constrained, allowing us to capture only the general trend of the power spectrum.
\begin{figure}
    \centering
    \includegraphics[width=0.95\linewidth]{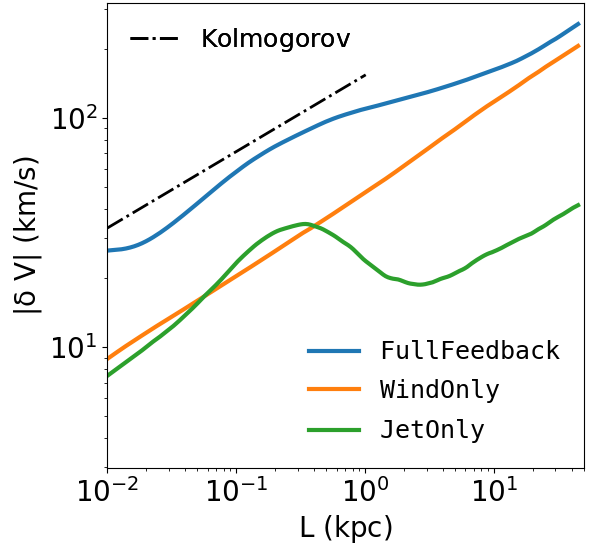}
    \caption{The power spectra of turbulent velocity plotted against physical length $L$ of the {\tt FullFeedback}, {\tt WindOnly} and {\tt JetOnly} simulations, superposed with the Kolmogorov slope of $L^{-5/3}$ (dash-dotted line). {\tt WindOnly} and {\tt FullFeedback} simulations are close to Kolmogorov slope within $10\%$ of accuracy, while the {\tt JetOnly} simulation shows a substantial deviation from the Kolmogorov slope, with a dip in the power spectrum at a few kpc.}
    \label{fig:TES}
\end{figure}
Fig.~\ref{fig:TES} illustrates the power spectra of turbulent velocity plotted versus physical length $L$ in the {\tt FullFeedback}, {\tt WindOnly}, and {\tt JetOnly} simulations. The Kolmogorov power spectrum of $\delta v \propto L^{1/3}$ is superimposed as a dash-dotted line. Among the three simulations, {\tt FullFeedback} preserves the largest power at all scales, consistent with the results of turbulent energy. The power spectrum of {\tt WindOnly} is close to the Kolmogorov slope within $5\%$ of accuracy, which is expected since the hot wind tends to introduce isotropic structure. In contrast, {\tt JetOnly} exhibits notable deviation from the Kolmogorov slope, with a dip on kpc scales. This dip, with a scale coinciding with the size of the cocoon-like structure observed in {\tt JetOnly}, could be due to the thermalization of the jet energy.  For {\tt FullFeedback}, although jets indeed introduce strong anisotropy, as shown in Fig.~\ref{fig:gasproperty}, this anisotropy is alleviated by the winds in terms of the shape of the power spectrum, leading to a power spectrum close to the Kolmogorov slope within $10\%$ of accuracy. This suggests that in the wind-jet interaction scenario, the winds wrapping around the jets help isotropize the turbulence and lead to a power spectrum consistent with the Kolmogorov slope up to $\sim 10$ -- $10^2$ kpc. Admittedly, the turbulence cascade in 2D simulations is not as realistic as in 3D simulations, and we will further investigate the power spectrum in 3D simulations in the future \citep{zhang2025macer3d}.

\begin{figure}
    \centering
    \includegraphics[width=0.95\linewidth]{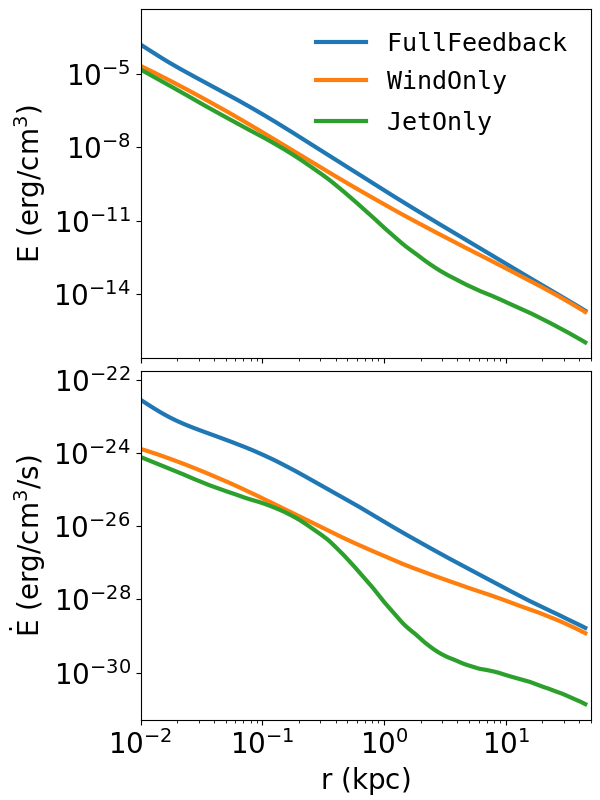}
    \caption{Profiles of the turbulent energy density (top) and energy dissipation rate (bottom) in the {\tt FullFeedback} (blue), {\tt WindOnly} (orange), and {\tt JetOnly} (green) simulations, averaged over the entire simulation time. The {\tt FullFeedback} simulation shows the highest turbulent energy and dissipation rates, followed by {\tt WindOnly} and then {\tt JetOnly}. The jet feedback alone is ineffective in generating turbulence and heating the gas beyond kpc scales.}
    \label{fig:e_turb}
\end{figure}
To quantify the turbulent energy dissipation, Fig.~\ref{fig:e_turb} shows the profiles of the turbulent energy density (top) and energy dissipation rate (bottom) in the {\tt FullFeedback} (blue), {\tt WindOnly} (orange), and {\tt JetOnly} (green) simulations, averaged over the entire simulation time. We note that calculating the turbulent energy dissipation rate is highly non-trivial, and we roughly estimate the characteristic turbulent energy and dissipation rate at each radius as the kinetic energy and turnover time of the largest eddy at that radius. Both turbulent energy and energy dissipation rate decline with radius due to the decreasing density and velocity dispersion. The {\tt FullFeedback} simulation shows the highest turbulent energy and dissipation rates, followed by {\tt WindOnly} and then {\tt JetOnly}. This suggests that the jet feedback alone is less effective in generating turbulence and heating the gas, in particular, at larger scales beyond $\sim 1\,\mathrm{kpc}$, due to the less collimated jet structure and the lack of interaction with the hot wind. The wind feedback alone is more effective at larger scales due to its isotropic kinetic energy injection compared to {\tt JetOnly}, but the combination of both feedback mechanisms is the most effective in generating turbulence and heating the gas.

\section{Discussions}
\label{sec:implications}

\subsection{AGN Feedback Efficiency and Turbulence Generation}
Whether AGN jets can effectively suppress star formation is a long-standing question. Previous numerical studies have shown that AGN jets can introduce turbulence and shocks into the ISM, but the efficiency of this process is still under debate \citep{2016Yanga,2018Guo}. At scales of clusters, AGN jets primarily heat the gas through weak shocks, while the heating rate might be insufficient to offset the radiative cooling of the gas \citep{2016Yanga}. Alternative approaches to maximize the heating efficiency of AGN jets include (i) increasing the effective interaction area of the jets, e.g., by increasing the jet opening angle or introducing jet precession \citep{2021Talbot,2023Husko}, and (ii) increasing the effective interaction time of the jets, e.g., by launching high-temperature, long cooling-time jets or by introducing cosmic rays that can propagate further than the jets \citep{2021Su}. This study provides a new perspective on this issue by considering the interaction between AGN jets and winds, which has been largely overlooked in previous studies. The coupling between AGN jets and winds can significantly enhance the efficiency of AGN feedback, as demonstrated by the {\tt FullFeedback} simulation.

\subsection{Comparison with Observations}
Observations provide important clues to the existence and effects of turbulence in galactic environments. Chandra observations have revealed that the turbulent dissipation in the ICM is sufficient to offset the radiative cooling locally \citep{2014Zhuravleva}. 

\citet{2020Li} have further shown that the velocity structure function of the H$\alpha$ filaments follows a $v \sim k^{-1/3}$ power law, which is consistent with the Kolmogorov-like power spectrum.
More recently, turbulence has been detected in the CGM of galaxies through both absorption \citep{qu2022cosmic,chen2023cosmic} and emission \citep{chen2023empirical,chen2024ensemble} lines, finding the power spectrum of the velocity field to be consistent with the Kolmogorov-like power spectrum as well. These impose strong constraints on the nature of the turbulence in the CGM. Our results are consistent with these observations, as the {\tt FullFeedback} simulation exhibits a power spectrum close to the Kolmogorov slope, while the {\tt JetOnly} simulation shows a substantial deviation from the Kolmogorov slope with much less turbulence. This suggests that the interaction between AGN jets and winds is a promising channel in generating turbulence with a power spectrum consistent with the observed turbulence in the ICM and CGM.

Although isolating the contribution of turbulent heating from other heating mechanisms (weak shocks, SN feedback, etc.) is challenging in our simulations, a rough estimate of the turbulent heating rate is $\sim 10^{-27}\,\mathrm{erg/cm^3/s}$ on scales of $\sim 10\,\mathrm{kpc}$ (Fig.~\ref{fig:e_turb}), consistent with observational constraints on turbulent heating rate in the CGM of elliptical galaxies \citep{chen2024ensemble}.  The {\tt WindOnly} simulation is still consistent with the observed turbulent heating rate, while the {\tt JetOnly} simulation is below the observed value by a few orders of magnitude, indicating that the jet feedback alone is unable to generate sufficient turbulence in the CGM. Future X-ray observations, such as eROSITA \citep{2012Merloni}, XRISM \citep{2020XRISM}, and HUBS \citep{2020HUBS,bregman2023scientific}, will be able to put forward more stringent constraints on the turbulent properties in the ICM and CGM, providing a direct test of our results on the properties of turbulence in these environments.

\subsection{Caveats and Future Directions}

While our simulations provide valuable insights into AGN jet-wind interaction and its role in regulating star formation, several important caveats warrant discussion. 

First, the 2D axisymmetric nature of our simulations inherently limits the development of turbulence and hydrodynamic instabilities. In particular, the KH instability can only develop in the meridional plane, while the parasitic instabilities that would normally cascade energy to smaller scales in three dimensions are suppressed. This geometric constraint likely underestimates the true turbulent energy cascade rate and may artificially enhance the coherence of large-scale structures. Additionally, 2D simulations cannot capture the full 3D topology of vortex tubes and the inverse cascade of magnetic energy that would be present in a fully magnetized medium. To address these limitations, we have recently developed a 3D version of the MACER framework \citep{zhang2025macer3d}, which will allow us to investigate the jet-wind interaction in a more realistic setting and quantify the impact of dimensionality on turbulence generation and energy dissipation.

Second, our isolated galaxy setup neglects the cosmological context, including cosmological inflows of gas from the intergalactic medium and the influence of the large-scale environment. In realistic scenarios, cosmological accretion can replenish the gas reservoir and alter the thermodynamic state of the CGM, potentially modifying the efficiency of AGN feedback. The absence of environmental effects, such as ram pressure stripping in cluster environments or tidal interactions with neighboring galaxies, may also affect the gas dynamics and star formation history. Future work incorporating cosmological simulations with sufficient resolution to capture both large-scale structure formation and small-scale AGN feedback processes will be essential for understanding the coevolution of galaxies and their supermassive black holes in a cosmological context.

Third, our purely hydrodynamic approach omits potentially crucial non-thermal physics. Magnetic fields, which are ubiquitous in galactic environments, can provide additional pressure support, modify the development of turbulence and turbulent mixing (e.g., \citealt{2019Ji,2023Zhao,das2023magnetic}), and enable the generation and propagation of cosmic rays \citep{2013Zweibel,2023Ruszkowski}. Magnetic fields may also potentially serve as an important feedback mechanism at cluster scales (e.g., \citealt{2024Cen}). Cosmic rays, accelerated at shocks and carried by jets, can provide an additional heating channel and may contribute significantly to the pressure support of the CGM \citep{ji2020properties,quataert2025cosmic}. Incorporating magnetic fields and cosmic rays in our framework represents an important avenue for future investigation, particularly given observational evidence for strong magnetic fields and non-thermal emission in jet-driven outflows.

Looking ahead, we plan to systematically investigate AGN jet feedback in 3D simulations with varying jet parameters, black hole spins, and galaxy properties to establish more robust scaling relations. We will also work toward incorporating our AGN feedback model into cosmological simulations to study the statistical properties of the galaxy population and the role of AGN feedback in shaping the observed galaxy scaling relations. Finally, the inclusion of magnetic fields and cosmic rays will allow us to explore the relative importance of thermal and non-thermal processes in mediating energy transfer from AGN outflows to the ISM and CGM.

\section{Conclusions}
\label{sec:conclusions}

In this study, we investigate the impact of AGN feedback, in particular, with the combined effects of the AGN winds and jets taken into consideration, on the evolution of a massive elliptical galaxy. Employing hydrodynamical simulations with self-consistent two-mode AGN feedback physics implemented and the Bondi radius resolved, we systematically examine the influence of AGN feedback on the evolutionary path of the host galaxy by controlling the activation and deactivation of AGN jets and winds and comparing the results regarding the gas thermal and kinematic properties, star formation, etc.

The most important finding of this work is that the AGN winds and jets are \emph{non-linearly} coupled; i.e., the outcomes of the coexistence of both AGN jets and winds (the {\tt FullFeedback} run) lead to far more efficient turbulence generation and gas heating than either feedback channel alone (the {\tt WindOnly} and {\tt JetOnly} runs). The key reason is that in the {\tt FullFeedback} run, the jet-wind velocity shear induces KH instabilities, which effectively produces turbulence with a Kolmogorov-like energy spectrum, leading to efficient heating at scales of $\sim 10\,\mathrm{kpc}$ and subsequent suppression of star formation. In contrast, the jets alone are much less collimated, forming a cocoon-like structure in the biconical regions, without producing sufficient turbulence to suppress star formation. The winds alone are more isotropic, with their effects on the gas dynamics and thermal properties falling between those observed in the {\tt FullFeedback} and {\tt JetOnly} cases.

The outcomes of the AGN wind-jet interaction have several important implications for the gas properties and the evolution of massive elliptical galaxies, as summarized below.

\begin{itemize}

  \item \emph{Jet collimation and velocity shears} In {\tt FullFeedback}, the AGN winds help collimate the jets, leading to strong velocity shear up to $\sim 10^3\,\mathrm{km/s}$ across the jets, winds and the ISM. In the absence of the AGN winds, the jets are less collimated and form low-volume-filling cocoon-like structures, producing the least amount of velocity shear at $\sim 10^2\,\mathrm{km/s}$. The cases with jets included, i.e., {\tt FullFeedback} and {\tt JetOnly}, produce strong anisotropy in the gas thermal and kinematic properties, while the {\tt WindOnly} case is more isotropic.
  
  \item \emph{Enhanced shear-induced turbulence} The velocity field analysis reveals that winds substantially enhance jet effectiveness by creating optimal conditions for KH instabilities. In {\tt FullFeedback}, the interaction between collimated jets and surrounding winds generates steep velocity gradients across multiple scales, whereas {\tt JetOnly} produces minimal shear due to rapid jet expansion. This fundamental difference in shear development explains why combined feedback generates strong turbulence and energy dissipation, with KH growth timescales comparable to local dynamical timescales -- a condition not achieved when either mechanism operates alone.
  
  \item \emph{Solenoidal vs. compressive} Within the central region of $r \lesssim 20\,\mathrm{kpc}$, the velocity fields in the {\tt FullFeedback} and {\tt WindOnly} simulations are dominated by solenoidal modes, where the ratio of solenoidal to compressive modes is higher in the {\tt FullFeedback} case. This suggests that the jet-wind interaction is more effective in producing turbulence than the jet or wind alone via shearing motions. In contrast, the {\tt JetOnly} simulation is dominated by compressive modes within the same region, indicating the existence of prevalent shocks and the lack of turbulence.

  \item \emph{High-entropy gas in the central region} Turbulence dissipation driven by the jet-wind interaction significantly suppresses cool gas formation and enhances the gas entropy within the central region of $r \lesssim 10\,\mathrm{kpc}$ in the {\tt FullFeedback} run, leading to a relatively flattened entropy profile. However, neither the {\tt WindOnly} nor the {\tt JetOnly} runs can maintain such a flattened profile. The {\tt FullFeedback} run also leads to lower density in the central region than the other two runs.
  
 \item \emph{Suppression of multiphase gas and star formation} By producing high-entropy, low-density gas in the central region, the {\tt FullFeedback} run effectively enhances the ratio of cooling time to free-fall time, minimizing the amount of gas falling below the cooling threshold of $t_\mathrm{cool}/t_\mathrm{ff} \sim 10$. This leads to a significant reduction in star formation, with a star formation rate of less than $10^{-3}\,\mathrm{M_\odot/yr}$. In contrast, the {\tt WindOnly} and {\tt JetOnly} runs contain more multiphase gas and 1 -- 2 orders of magnitude higher star formation rates. We note that this suppression is not due to higher total AGN energy output; on the contrary, {\tt JetOnly} injects 1 -- 2 orders of magnitude more energy than the other two runs yet remains the least effective at suppressing star formation.

\end{itemize}

Our results suggest that the AGN jets and winds are non-linearly coupled, and the jet-wind interaction is crucial for the efficient generation of turbulence and heating in the central region of massive elliptical galaxies. These findings provide new insights into the complex interplay between AGN feedback, turbulence, and the regulation of star formation in massive elliptical galaxies. Admittedly, certain caveats exist in our simulations, such as the 2D nature of the simulations. Recently, \citet{zhang2025macer3d} have integrated AGN feedback within the broader context of galaxy formation and galactic feedback. Ongoing work is underway to investigate the synergistic effects of AGN jets and winds in fully 3D simulations: one approach isolates the interaction effects of jets and winds in idealized studies with adaptive mesh refinement.

\begin{acknowledgments}
\emph{Acknowledgments} We thank the anonymous referee for constructive comments that improved the paper. We thank H. W. Chen, L. C. Ho, and Z. Qu for stimulating discussions. Authors are supported by the National Key R\&D Program of China No. 2023YFB3002502, the Natural Science Foundation of China (grants 12133008, 12192220, 12192223, and 12361161601), and the China Manned Space Program (grants CMS-CSST-2025-A08 and CMS-CSST-2025-A10). This work was performed in part at the Aspen Center for Physics, which is supported by National Science Foundation grant PHY-2210452. Numerical calculations were run on the CFFF platform of Fudan University, the supercomputing system in the Supercomputing Center of Wuhan University, and the High Performance Computing Resource in the Core Facility for Advanced Research Computing at Shanghai Astronomical Observatory. We have made use of NASA's Astrophysics Data System.
\end{acknowledgments}

\software{{\small Matplotlib} \citep{hunter2007matplotlib},
          {\small NumPy} \citep{2020NumPy-Array},
          {\small SciPy} \citep{2020SciPy-NMeth}, 
          {\small yt} \citep{Turk2010,turk2024introducing}
          }

\appendix
\section{Convergence Tests}
\subsection{Resolution Tests}

We conducted resolution tests to confirm numerical convergence of the turbulence statistics. Specifically, we performed a high-resolution run, {\tt FullFeedbackHR} ($512 \times 64$), with double the radial resolution of our fiducial run, {\tt FullFeedback} ($240 \times 64$). Due to computational cost, this high-resolution run was only evolved for 1.5 Gyr. The polar resolution was kept unchanged, ensuring that the smallest resolved scale remains the same for power spectrum calculations.

\begin{figure*}
    \centering
    \includegraphics[height=0.25\textheight]{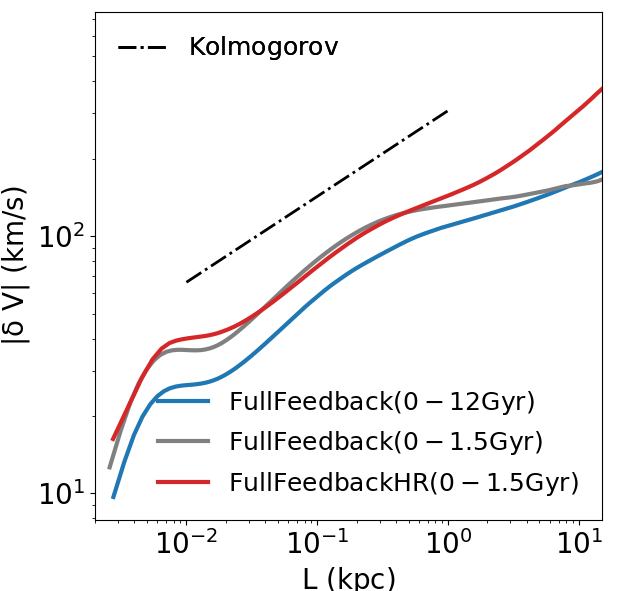}\hspace{0.02\textwidth}%
    \includegraphics[height=0.25\textheight]{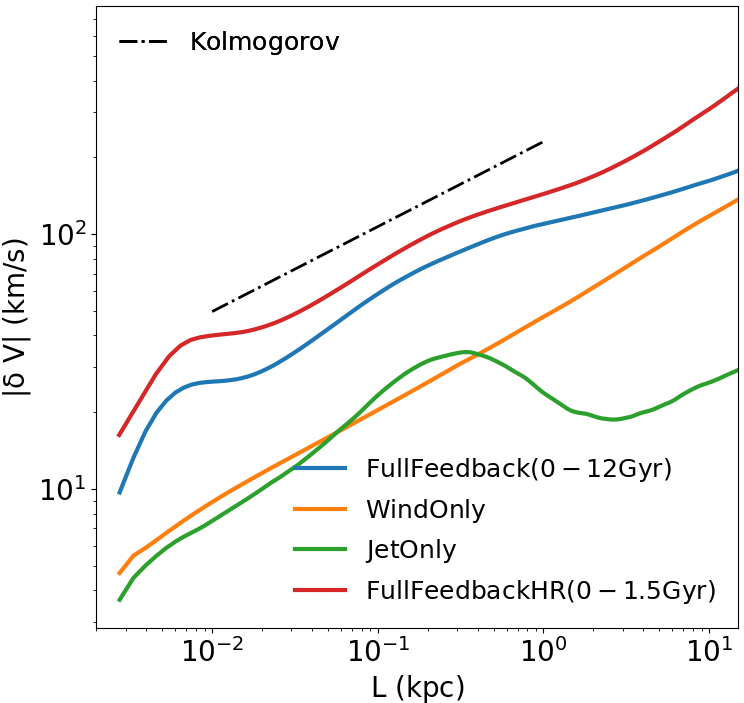} 
    \caption{Left panel: velocity dispersion profiles for the fiducial-resolution run ({\tt FullFeedback}, blue; $240\times 64$ grid) and the high-resolution run ({\tt FullFeedbackHR}, red; $512 \times 64$ grid). The time-averaged profile from 0–1.5 Gyr for the fiducial case is also shown (gray). Within 10 kpc, the profiles are nearly identical, indicating convergence of the turbulence statistics in the main star-forming region. At larger scales, the high-resolution run shows slightly higher velocity dispersion while maintaining a Kolmogorov-like slope.
    Right panel: comparison among different feedback modes ({\tt JetOnly}, {\tt WindOnly}, {\tt FullFeedback}) and {\tt FullFeedbackHR}.  The slightly higher amplitude in {\tt FullFeedbackHR} reflects stronger AGN activity at early times.}
    \label{fig:con}
\end{figure*}

As shown in Figure~\ref{fig:con}, within the main star-forming region ($< 1$~kpc), the velocity dispersion is nearly identical to that of the fiducial run over the first 1.5 Gyr. At larger scales ($> 1$~kpc), the high-resolution run exhibits slightly higher velocity dispersion while retaining a Kolmogorov-like scaling. This enhancement is attributed to reduced numerical dissipation, allowing more large-scale eddies and kinetic energy to be resolved. Importantly, these differences do not affect our main conclusions regarding the relative effectiveness of different feedback modes, indicating that the results are well converged in the region where AGN feedback regulates star formation.

\bibliographystyle{aasjournal}
\bibliography{main}

\end{document}